\documentclass[12pt]{article}  
 
\usepackage{latexsym}  %WORKS!!!!!!!!!!!!!!!!!!!!!!
\usepackage{quotes}
    
\usepackage{tikz}

\usepackage{booktabs}

\usepackage{slashed}
\usepackage{float}
\usepackage{pgfplots} 

\pgfmathsetmacro{\range}{4}
\pgfmathsetmacro{\gridgreylevel}{30}
\pgfmathsetmacro{\margin}{0.25}

\usepackage{latexsym}
\usepackage{graphicx}
\usepackage{amsfonts}
\usepackage{amssymb}
\usepackage{amsmath}
\usepackage{hyperref}

\usepackage{multirow}
\def\be{\begin{equation}}
\def\ee{\end{equation}}
\def\ba{\begin{eqnarray}}
\def\ea{\end{eqnarray}}

\newcommand{\beq}{\begin{equation}}
\newcommand{\eeq}[1]{\label{#1}\end{equation}}
\newcommand{\bea}{\begin{eqnarray}}
\newcommand{\eea}[1]{\label{#1}\end{eqnarray}}

\newcommand{\sst}[1]{{\scriptscriptstyle #1}}

\newcommand{\ft}[2]{\tfrac{#1}{#2}}

\def\0{{\sst{(0)}}}
\def\1{{\sst{(1)}}}
\def\2{{\sst{(2)}}}
\def\3{{\sst{(3)}}}
\def\4{{\sst{(4)}}}
\def\5{{\sst{(5)}}}
\def\6{{\sst{(6)}}}
\def\7{{\sst{(7)}}}

\def\CP{{{\mathbb C}{\mathbb P}}}

\pgfplotsset{compat=1.17}
\begin{document}

\begin{flushright}
\hfill {Imperial/TP/2025/mjd/1 \ \ \  MI-HET-851}\\
\end{flushright}

\vspace{20mm}

\begin{center}

{\bf Kaluza-Klein  Supergravity 2025*}
\vspace*{0.37truein}
\vspace*{0.15truein}

M.~J.~Duff\\
%\footnote{m.duff@imperial.ac.uk} \\
\vspace*{0.15truein}
{\it Blackett Laboratory, Imperial College London} \\
{\it Prince Consort Road, London SW7 2AZ} 
\vspace*{0.37truein}

B.~E.~W. Nilsson\\
%\footnote{tfebn@chalmers.se} \\
\vspace*{0.15truein}
{\it Department of Physics, Chalmers University of Technology} \\
{\it SE-412 96 G\"oteburg, Sweden} 
\vspace*{0.37truein}

C.~N.~Pope\\
%\footnote{pope@physics.tamu.edu} \\
\vspace*{0.15truein}
{\it George P and Cynthia Woods Mitchell Institute for Fundamental Physics 
and}\\
{\it Astronomy, Texas A\&M University, College Station, TX 77843 USA} 
\vspace*{0.15truein}

{\it DAMTP, Centre for Mathematical Sciences,
 Cambridge University,\\  Wilberforce Road, Cambridge CB3 OWA, UK}

%\end{center}

\baselineskip=10pt
\bigskip

\vspace{20pt}

\abstract

\noindent  We recall our work in the 1980s taking seriously the maximal eleven dimensions  of   supergravity, 
in particular the round, left squashed and right squashed $S^7$ compactifications to $D=4$ yielding $\mathcal N=8$,  
$\mathcal N=1$ and  $\mathcal N=0$, respectively. This involved Kaluza-Kein techniques that have found wider applications 
such as spontaneous compactification, holonomy and supersymmetry, topology versus geometry, squashing and Higgs, vacuum stability and 
consistent truncations.

\vspace{25pt}

\noindent *Invited contribution to the book {\it Half a Century of Supergravity}, eds. A. Ceresole and G. Dall’Agata. \end{center}
\newpage

\tableofcontents

\newpage

%%%%%%%%%%%%%%%%%%%%%%%%%%%%%%%%%%%%%%%%%%%%%%%%%%%%%%%%%%%%%%%%%%%%%%%%%%%%%

\section{Why Kaluza-Klein?}

Forty years ago our Physics Reports \cite{dunipo} began:

``We do not yet know whether supergravity \cite{Freedman:1976xh, Deser:1976eh, Fayet:1976cr,vanN} is a theory of the real world, nor whether the
real world has more than four dimensions as demanded by Kaluza-Klein theories \cite{Kaluza,klein1,klein2}. However,
our research in this area has convinced us that the only way to do supergravity is via Kaluza-Klein and
that the only viable Kaluza-Klein theory is supergravity.'' If we swap {\it is} for {\it has as its low-energy limit}, we think this was  a fairly accurate prophecy \cite{dunipo}.

 Noting that that the field equations of ${\cal N} = 1$ supergravity in $D = 11$
dimensions admit of vacuum solutions corresponding to AdS$_4 \times  S^7$, and noting that the round $S^7$ admits 8 Killing spinors and that its isometry group is $SO(8)$,  this gives rise via a Kaluza-Klein mechanism to an effective $D= 4 $ theory with  ${\cal N} = 8$ supersymmetry, local $SO(8)$ invariance with coupling $e$ and cosmological constant  $\Lambda$ with $G \Lambda \sim - e^2$
where $G$ is Newton's constant \cite{dufpopKK}.  
Noting further that the massless sector coincides  with the ${\cal N}=8$ gauged supergravity of de Wit and Nicolai \cite{deWit:1982bul}, we conjectured in 1982 (in the face of much opposition) that omission of the massive states provided a consistent  truncation \cite{dufpopKK}. The conjecture was not completely proved until 2012 \cite{NP}. It was these observations
which first aroused our interest in Kaluza-Klein theories and provided the stimulus for our subsequent
research.  Other sources of inspiration included \cite{Wittensearch,Salam:1981xd}.  Furthermore, we noted that $S^7$ admits a second Einstein metric, the squashed seven-sphere with ${\cal N}=1 $ and isometry $Sp_2 \times Sp_1$ \cite{Awada:1982pk}. A reversal of orientation (skew whiffing) leads to  a different compactification with 
the same isometry, but ${\cal N}=0$ \cite{Duff:1983ajq}. Notwithstanding  the absence of supersymmetry, it is 
classically stable \cite{Duff:1984sv}.
 It was to the seven-sphere, therefore, that the bulk of our Physics Report was devoted.

However, the example of $S^7$ revealed novel results and techniques that are valid in a wider class of Kaluza-Klein theories which we shall also describe. 
%%%%%%%%%%%%%%%%%%%%%%%%%%%%%%%%%%%%%%%%%%%%%%%%
\section{Spontaneous compactification}
%%%%%%%%%%%%%%%%%%%%%%%%%%%%%%%%%%%%%%%%%%%%%%%%
Soon after Nahm \cite{Nahm:1977tg} pointed out that $D=11 $ is the maximum dimension for supergravity, Cremmer, Julia and Scherk \cite{Cremmerjuliascherk} were able to construct the corresponding Lagrangian, but this was regarded as a mathematical curiosity
with no physical significance. A major goal  was to derive (in $D=4$) the extended supergravities, especially the one with
maximal ${\cal N}=8$. Cremmer and Julia \cite{Cremmerjulia} were able to do this using {\it Dimensional Reduction} which simply means demanding that all fields be independent of the extra 7 coordinates.

Gradually, however, theorists began to insist on {\it spontaneous compactification}, that is to look for stable ground state solutions of the field equations for which the metric describes a product manifold
$M_4 \times X^n$ where $M_4$ is four dimensional spacetime with the usual signature $(-,  +, +, +) $ and $X^n$ is a compact {\it internal} space with Euclidean signature $(+,+,+,...)$. This would yield Yang-Mills fields in $D=4$ with gauge group $G$, where $G$ is the isometry group of $X^n$, with coupling constant $e$ where
\be
e^2 \sim Gm^2,
\label{e}
\ee
 and $m^{-1}$ is the radius of the compact manifold. For better or worse such compactifications also typically involved a cosmological constant $\Lambda$
\be
\Lambda \sim -m^2.
\label{lambda}
\ee
 An important example was the Freund-Rubin compactification described below. Indeed one of the observation that triggered our combining Kaluza-Klein and supergravity was that combining (\ref{e}) and (\ref{lambda})
 we find 
 \be
 G\Lambda \sim -e^2,
 \label{G}
 \ee
 as in gauged $\mathcal N=8$ supergravity.

 The unique $D=11,\, {\cal N}=1$ supermultiplet is
comprised of a graviton $g_{MN}$, a gravitino $\Psi_M$ and $3$-form
gauge field $A_{MNP}$, where $M=0, 1, \ldots 10$, with $44$, $128$ and
$84$ physical degrees of
freedom, respectively. 
The transformation  rule of the gravitino reduces in a
purely bosonic background to
\begin{equation}
\delta \Psi_{M}={\tilde D}_{M} \epsilon,
\end{equation}
where the parameter $\epsilon$ is a 32-component anticommuting Majorana spinor,
and where
\begin{equation}
{\tilde
D}_{M}=D_{M}-
\frac{i}{144}
(\Gamma^{NPQR}{}_{M}
+8\Gamma^{PQR}
\delta^{N}{}_{M})F_{NPQR}.
\label{supercovariant}
\end{equation}
Here $D_{M}$ is the usual Riemannian covariant derivative involving the usual
Levi-Civita connection $\omega_{M}$
\begin{equation}
 D_{M}=\partial_{M}-\frac{1}{4}\omega_{M}{}^{AB}\Gamma_{AB},
\end{equation}
$\Gamma^{A}$ are the $D=11$ Dirac matrices and $F=dA$.  The bosonic field equations are
\begin{equation}
R_{MN}-\frac{1}{2}g_{MN}R=
\frac{1}{3}\left(F_{MPQR}F_{N}{}^{PQR}-\frac{1}{8}g_{MN}F^{PQRS}F_{PQRS}\right),
\end{equation}
and
\begin{equation}
d*F+F \wedge F=0.
\end{equation}
The Freund-Rubin ansatz \cite{Freundrubin} is
\begin{equation}
F_{\mu\nu\rho\sigma}=3m\epsilon_{\mu\nu\rho\sigma},
\end{equation}
where $\mu=0,1,2,3$ and $m$ is a constant with the dimensions of mass.
 This effects a
  spontaneous compactification from $D=11$
to $D=4$, yielding the product of a four-dimensional spacetime with negative
curvature
\begin{equation}
R_{\mu\nu}=-12m^{2}g_{\mu\nu},
\end{equation}
and a seven-dimensional internal space of positive curvature
\begin{equation}
R_{mn}=6m^{2}g_{mn},
\end{equation}
where $m=1,2,\ldots 7$.  Accordingly, the supercovariant derivative also splits
as
\begin{equation}
{\tilde D}_{\mu}= D_{\mu}+m\gamma_{\mu}\gamma_{5},
\end{equation}
and
\begin{equation}
\label{covariant}
{\tilde D}_{m}= D_{m}-\frac{1}{2}m\Gamma_{m}.
\end{equation}
If we choose the spacetime to be maximally symmetric but leave the internal
space
$X^{7}$ arbitrary, we are led to the $D=11$ geometry AdS$_{4} \times X^{7}$.
 The first example was provided by the choice $X^{7}=$ round $S^{7}$
 \cite{dufpopKK, dunipo} which is maximally
 supersymmetric\footnote{The first Ricci flat ($m=0$) example of a
 compactification of $D=11$ supergravity was provided by the choice
 $X^{7}=T^{7}$ \cite{Cremmerjulia} which is also maximally
 supersymmetric.}.  The next example was the round $S^{7}$ with
 parallelizing torsion \cite{Englert} which preserves no supersymmetry.
 However, it was also of interest to look for something in between, and
 this is where holonomy comes to the fore.

%\section{Killing spinors, holonomy and supersymmetry}
%%%%%%%%%%%%%%%%%%%%%%%%%%%%%%%%%%%%%%%%%%%%%
\section{Holonomy, Killing spinors and supersymmetry}
%%%%%%%%%%%%%%%%%%%%%%%%%%%%%%%%%%%%%
 Crucial to the whole programme of Kaluza-Klein supergravity are the notions of Killing spinors \cite{Duffnilssonpopecomp,dunipo} and holonomy groups \cite{Awada:1982pk, Duff:1983ajq,dunipo}.  In 1981, Witten laid down the criterion for spacetime supersymmetry in Kaluza-Klein theory  \cite{Wittensearch}.  The number of spacetime supersymmetries is given by the number of covariantly constant spinors on the compactifying manifold.  To see 
this, we look for vacuum solutions of the field equations for which the
the gravitino field $\Psi$ vanishes.  In order that the vacuum be
supersymmetric, therefore, it is necessary that the gravitino remains
zero when we perform a supersymmetry transformation and hence that the
background supports spinors $\epsilon$ satisfying 
\begin{equation}
{\tilde D}_{M}\,\epsilon=0.
\end{equation}
In the case of a product manifold, this reduces to
\begin{equation}
{\tilde D}_{\mu}\,\epsilon(x) = 0,
\end{equation}
and
\begin{equation}
{\tilde D}_{m}\,\eta(y)= 0,
\label{Killing}
\end{equation}
where $\epsilon(x)$ is a 4-component anticommuting spinor and
$\eta(y)$ is an 8-component commuting spinor.  The first equation is satisfied
automatically with our choice of AdS$_{4}$ spacetime and hence the number
of $D=4$ supersymmetries, $0\leq {\cal N} \leq 8$, devolves upon the number of {\it
Killing spinors} on $X^{7}$.  They satisfy the integrability condition
\begin{equation}
[{\tilde D}_{m}, {\tilde D}_{n}] \eta= -\frac{1}{4}C_{mn}{}^{ab}\Gamma_{ab}\eta=0,
\label{integrability}
\end{equation}
where $C_{mn}{}^{ab}$ is the Weyl tensor.

As pointed out in \cite{Awadaduffpope} covariantly constant spinors, or {\it Killing spinors} are, in their turn, related to the holonomy group of the corresponding
connection. It was well known that the number of massless gauge bosons was determined by the isometry group of the compactifying manifold, but it turned out
to be the holonomy group that determined the number of massless gravitinos. The subgroup of $Spin(7)$ generated by these linear combinations of $Spin(7)$ generators
$\Gamma_{ab}$ corresponds with the {\it holonomy } group ${\it H}$.

The first non-trivial example was provided in 1982 \cite{Awada:1982pk} by compactifying $D = 11$
supergravity on the squashed $S^7$, an Einstein space whose whose $G_2$ holonomy
yields $ {\cal N} = 1$ in $D = 4$. Here  we are referring to the holonomy of the supercovariant  derivative 
${\tilde D}_{m}$ that appears in (\ref{covariant}). This differs from the Riemannian 
or {\it Levi-Civita} derivative $D_m$ that you find in the maths textbooks. Similarly, the round $S^7$ has trivial holonomy and hence yields the
maximum ${\cal N} = 8$ supersymmetry \cite{Duffnilssonpopecomp}. 

From the Riemannian perspective,  $G_2$ corresponds to the {\it weak holonomy} of $S^7$.  A 7-dimensional Einstein manifold with $R_{mn} = 6m^2g_{mn}$  has weak holonomy $G_2 $ if it admits a 3-form A obeying
\be
dA = 4m * A. 
\ee
 That such a 3-form exists on the squashed $S^7$ can be proved by invoking the single
(constant) Killing spinor $\eta$. The required 3-form is then given by 
\cite{Duff:1983ajq}
\be
A_{mnp}={\bar \eta} \Gamma_{mnp} \eta.
\ee

Although the phenomenologically desirable
${\mathcal N=1}$ supersymmetry  and non-abelian gauge groups appear in four dimensions,
the resulting theory was not realistic, being vectorlike with gauge group $Sp_2 \times Sp_1 $ and
living on AdS$_4$. It nevertheless provided valuable insight into the workings of
modern Kaluza-Klein theories. Forty years later, $G_2$ manifolds continue to play
an important role in $D = 11$ M-theory for the same ${\cal N} = 1$ reason. But the full
M-theory, as opposed to its low energy limit of $D = 11$ supergravity, admits the
possibility of singular $G_2$ compactifications which can yield chiral $({\cal N} = 1, D = 4)$
models living in Minkowski space and with realistic gauge groups \cite{Acharyadenef, Atiyahwitten}.

Owing to this generalized connection, vacua with $m\neq 0$ present subtleties
and novelties not present in the $m=0$ case, for
example the
phenomenon of {\it skew-whiffing}{
\cite{Duff:1983ajq,dunipo}.  For each
Freund-Rubin compactification, one may obtain another by reversing the
orientation of $X^{7}$.  The two may be distinguished by the labels
{\it left} and {\it right}.  An equivalent way to obtain such vacua is
to keep the orientation fixed but to make the replacement
$m\rightarrow -m$.  So the covariant derivative (\ref{covariant}), and
hence the condition for a Killing spinor, changes but the
integrability condition (\ref{integrability}) remains the same.  With
the exception of the round $S^{7}$, where both orientations give
${\cal N}=8$, at most one orientation can have ${\cal N} \geq 0$.  This is the {\it
skew-whiffing theorem}.  A corollary is that other {\it symmetric
spaces}, which necessarily admit an orientation-reversing isometry,
can have no supersymmetries.  Examples are provided by products of
round spheres.  Of course, skew-whiffing is not the only way to obtain
vacua with less than maximal supersymmetry.  A summary of known
$X^{7}$, their supersymmetries and stability properties is given in
\cite{dunipo}.  Note, however, that skew-whiffed vacua
are automatically stable at the classical level since skew-whiffing
affects only the spin $3/2$, $1/2$ and $0^{-}$ towers in the
Kaluza-Klein spectrum, whereas the criterion for classical stability
involves only the $0^{+}$ tower \cite{Duff:1984sv, dunipo}.

Once again the squashed $S^{7}$ provided the
first non-trivial example: the left squashed $S^{7}$ has ${\cal N}=1$ but the right
squashed $S^{7}$ has ${\cal N}=0$.  Interestingly enough, this means that
setting the suitably normalized 3-form  equal to the $D=11$
supergravity 3-form provides a solution to the field equations, but
only in the right squashed case.  This solution is called the {\it
right squashed $S^{7}$ with torsion} \cite{Duff:1983ajq}  since
$A_{mnp}$ may be interpreted as a Ricci-flattening torsion \cite{Englert}.  Other
examples were provided by the left squashed $N(1,1)$ spaces
\cite{pagpop3}, one of which has ${\cal N}=3$ and the other ${\cal N}=1$, while the
right squashed counterparts both have ${\cal N}=0$.

All this presents a dilemma.  If the Killing spinor
condition changes but the integrability condition does not, how does
one give a holonomic interpretation to the different supersymmetries?
Indeed ${\cal N}=3$ is not allowed by the usual rules.  The answer to this
question may be found in a paper \cite{Castellani} written before we
knew about skew-whiffing.  The authors note that in (\ref{covariant}),
the $SO(7)$ generators $\Gamma_{ab}$, augmented by presence of
$\Gamma_{a}$, together close on $SO(8)$.  Hence one may introduce a
generalized holonomy group ${\cal H}_{gen}\subset SO(8)$ and ask how
the $8$ of $SO(8)$ decomposes under ${\cal H}_{gen}$.  In the case of
the left squashed $S^{7}$, ${\cal H}_{gen}= SO(7)^{-}$, $8 \rightarrow
1+7$ and ${\cal N}=1$, but for the right squashed $S^{7}$, ${\cal H}_{gen}=
SO(7)^{+}$, $8 \rightarrow 8$ and ${\cal N}=0$. For more on generalized holonomy, see
\cite{Duff:2003ec,Hullhol}.

If the space is not simply connected there may be further global obstructions to the existence of unbroken supersymmetries. For example, solutions of the form $T^7/\Gamma$ and $S^7/\Gamma$, where $\Gamma$ is a discrete group, admit fewer than 8 Killing spinors. A prominent example of the latter is provided by the 
ABJM theory \cite{abjm}.

%%%%%%%%%%%%%%%%%%%%%%%%%%%%%%%%%%%%%%%%%
\section{Topology}
%%%%%%%%%%%%%%%%%%%%%%%%%%%%%%%%%%%%%%%%%
In \cite{Duffnilssonpopecomp} it was pointed out that Euclidean signature field configurations
and their topological properties (Betti numbers, Euler numbers, Pontryagin numbers, holonomy, index theorems etc) which feature in gauge
and gravitational instanton physics can lead a second life as internal
manifolds $X^n $ appearing in the compactification of the $n$ extra dimensions in Lorentzian signature Kaluza-Klein 
theory $ M_D = M_{D-n}  \times X^n$. 

The first non-trivial example of this kind was provided by the $K3$ manifold\footnote{K3 had already entered the physics literature through gravitational instantons \cite{Hawkingpopesymm} and non-linear sigma models \cite {Alvarez-Gaume}.}
 which is four-dimensional, self-dual, and Ricci flat without isometries \cite{Duffnilssonpopecomp}. It's
$SU(2)$ holonomy yields half the maximum supersymmetry, e.g., $({\cal N} = 2, D = 6) $ for $ K3$ and $({\cal N}=4,D=4) $ for $K3 \times T^2$ starting from 
$({\cal N} = 1, D = 10)$\footnote{In \cite{Duff:1986cx} it was pointed out that $D=11$ supergravity on $R^{10-n} \times K3 \times T^{n-3}$
 and the $D=10$ heterotic string on $R^{10-n} \times T^n$ not only have the same 
supersymmetry but also the same moduli space of vacua, namely $SO(16+n,n)/(SO(16+n) \times SO(n))$. }.  For the first time, the
Kaluza-Klein particle spectrum was dictated by the topology rather than the geometry of the compactifying manifold. It was thus a forerunner of the
six-dimensional Ricci-flat Calabi-Yau compactifications \cite{cahostwi}, whose $SU(3) $ holonomy
yields $ ({\cal N} = 1, D = 4)$ starting from $ ({\cal N} = 1, D = 10)$, and the seven-dimensional Ricci-flat Joyce compactifications \cite{Joyce1,Joyce2} whose  $G_2$ holonomy yields $({\cal N}=1,D=4) $ starting from 
$({\cal N}=1,D=11)$. $K3$ continues to feature prominently in M-theory and its dualities. 

Concerning Calabi-Yau, in 1983 the three authors were visiting Steve Weinberg's group in Texas.  We had recently realized the role played by holonomy in fixing the fraction of supersymmetry that survives compactification. Having noted that $({\cal N}=1, D=10)$ supergravity on $K3 \times T^2$ yields $({\cal N}=2, D=4)$, we then asked ourselves whether there was an $SU(3)$ analogue yielding $ ({\cal N}=1,D=4)$. However, this took us beyond our mathematical competence, so we went over to the Math Department and spoke to a well-known geometer who was there at the time. Now we do not want to blame him, because perhaps we physicists did not articulate clearly what we wanted, but in any event we came away with the impression that there was no known $SU(3)$ analogue.

Hence the statement in our paper  \cite{Duffnilssonpopecomp}  ''we do not know of any solutions... with ${\it H}=SU(3)$''.

%%%%%%%%%%%%%%\section{Why Kaluza-Klein?}

%%%%%%%%%%%%%%%%%%%%%%%%%%%%%%%%
\section{Squashing, Higgs and space invaders}
%%%%%%%%%%%%%%%%%%%%%%%%%%%%%%%%%%%%%%%%%%%%%%
It was realized already in the early 1980s \cite{Duff:1982ev, Duff:1983ajq,dunipo}
that the process in which the AdS$_4$ supergravity theory with eight supersymmetries based on the round $S^7$ compactification is turned into a similar theory based on the left squashed $S^7$ involves  various kinds of 
Higgs effects\footnote{That deforming the $S^7$ and torsion corresponded to non-zero vevs for scalars and pseudoscalars was recognised in  \cite{Duff:1982ev}. But it was \cite{page} that identified the  $Sp_2 \times Sp_1$ singlet in the 300 of $SO(8)$ as the source of squashing in the Squashed $S^7$.}.
In particular it was noticed that the eight massless spin-3/2 Rarita-Schwinger fields
responsible for the supersymmetries in  AdS$_4$ behave in an unexpected and interesting way. Instead of being subjected to a standard super-Higgs effect where one of the eight supersymmetries
in the round vacuum survives the squashing one finds that the  eight spin-3/2 fields in a spinor  irrep of $SO(8)$ all become massive when the isometry group breaks
from $SO(8)$ to $Sp_2 \times Sp_1$. This part of the story therefore constitutes a slightly non-standard super-Higgs effect. In this process the corresponding Goldstinos must arise in the breaking of the round AdS$_4$ spectrum only to have disappeared in the squashed vacuum since they are eaten in the process. This has recently been verified to be compatible with details of the two spectra involved here \cite{Nilsson:2018lof, Nilsson:2023ctq, Karlsson:2023dnl}.

The issue of where the single massless Rarita-Schwinger field in the squashed AdS$_4$ vacuum is coming from was on the other hand established in the 1980s.  It was referred to as  the  {\it space invaders scenario} \cite{dunipo}
since here a  singlet massive Rarita-Schwinger field, which is produced in the breaking of a particular round sphere irrep, zooms down to become massless. It was checked using the results in \cite{Nilsson:1983ru}  that this singlet could be traced under squashing to the new vacuum  and that
it indeed becomes the massless field needed to explain the single supersymmetry in this vacuum. Due to the fact that the 
breaking pattern is different when going to the right squashed vacuum it is easy to see that this does not happen there.

What was not investigated in the early literature on this subject was what happens to the singlet Goldstino that necessarily must be spat out in this de-Higgsing of the massive 
spin-3/2 singlet field. A possible explanation was suggested  in \cite{Nilsson:2018lof}
and to some extent verified in  \cite{Nilsson:2023ctq} by an explicit construction of all the squashed $S^7$ singlet mode functions\footnote{We emphasize here that this and the following discussions will often refer to modes and fields. These must be kept apart: {\it modes} refer to Fourier modes used in the expansions of tensor fields on $S^7$ while {\it fields} always are objects in the  effective low energy field theory in AdS$_4$. Fields in AdS$_4$ are in unitary irreps of $SO(2,3)$ denoted $D(E_0, s)$ where $E_0$ is the energy value of the lowest weight in the weight diagram  and $s$ its spin \cite{Nicolai:1984hb}. Modes and fields   are, however,  connected through relations like $E_0=E_0(M^2)$, $M^2=M^2(\Delta_p)$ and 
$\Delta_p=\Delta_p(C_G)$ where $C_G$ is the Casimir of the isometry group $G$. For details, see \cite{dunipo} and for the squashed $S^7$ case the more recent and complete results in \cite{Karlsson:2021oxd, Karlsson:2023dnl}. Partial results can also be found in \cite{dubmalsam}.}. What seems to be happening is that the singlet spin-3/2 mode that exists in the squashed $S^7$ spectrum
but not in the round one gives rise to a spin-1/2 field that belongs to a  Wess-Zumino multiplet in the $\mathcal N =1$ left squashed theory but generates  a singlet fermionic massive field in the right vacuum.
The fact that this spin-3/2 mode is not present in the round vacuum seems to 
indicate that an AdS$_4$ field must be eating it when returning to the 
round theory.
 A plausible explanation is that the massless 
spin-3/2 field eats it when going from the left vacuum to the round one  and a singleton when 
going back from the right vacuum. 
%This was also verified in \cite{Nilsson:}.

To summarize the situation described above in a bit more explicit terms we recall  from
 \cite{Nilsson:2023ctq} that the  singlet spinorial modes in the different vacua are:
One spin-1/2 and no spin-3/2 in the round $S^7$ $SO(8)$ spectrum broken  down to  the squashed isometry 
$Sp_2\times Sp_1$, and 
in the left and right squashed vacua, one spin-1/2 and one spin-3/2.

The modes are of course the same on the left and right spheres but their operator eigenvalues 
differ, when letting $m\rightarrow -m$,  which will affect the masses of the corresponding spin-1/2 and 3/2 fields in the AdS$_4$ theory.
To explain this mode-field connection we recall that, via the mass-operator relations (see, e.g., \cite{dunipo}),   spin-3/2 modes generate only 
spin-1/2 fields in AdS$_4$ while spin-1/2 modes give rise to both spin-1/2 and 3/2 fields in AdS$_4$. The modes listed above therefore  imply the following AdS$_4$ singlet field content in the various cases  \cite{Nilsson:2023ctq}:\\

\noindent {\bf Round vacuum}: One massive spin-3/2 field and one massive spin-1/2 field both compatible with the $\mathcal N = 8$ 
supersymmetric $SO(8)$  spectrum broken down to $Sp_2\times Sp_1$.\\
{\bf Left squashed vacuum}: One massless spin-3/2 field and two massive spin-1/2 fermions all compatible with $\mathcal N = 1$ supersymmetry.\\
{\bf Right squashed vacuum}: One massive spin-3/2 field, together with one massive spin-1/2 and one spin-1/2 singleton field.
\medskip

The only way to make these three cases compatible with each other seems to be to introduce a new kind of Higgs effect whereby a singleton can absorb a massive field of the same spin
to become a new massive field. This way both the left and right squashed  singlet spectra can be related  to the round one, and to each other. Note that in both squashed cases a spin-1/2 field must be eaten
since it does not appear in the round spectrum.  In the left case it is the massless spin-3/2 field that is doing the eating and in the right case it is the fermionic singleton.

As argued in \cite{Karlsson:2023dnl}
we can use the results of \cite{Nilsson:1983ru} to follow the massless spin-1/2 field in the irrep $(0,0;2)$, that is the superpartner of the gauge field in this irrep in the left squashed vacuum, back to the round
$S^7$. It turns out that it ends up as a singleton in the round case.

%%%%%%%%%%%%%%%%%%%%%%%%%%%%%%%

\section{The criterion for vacuum stability}

  The question of whether or not a particular vacuum solution in supergravity
is stable can be addressed at various levels.  The most basic one concerns
the classical stability of the solution.  Since the equations of motion are
non-linear, addressing even this classical question can be quite tricky, but
at the linearised level, it becomes relatively straightforward.  Thus one
can linearise the equations of motion around the chosen ``vacuum,'' i.e. the
background solution, and study the modes describing  the linearised
fluctuations.  In the Kaluza-Klein context, this amounts to studying the
spectra of all towers of lower-dimensional fields.   The signal for 
a classical instability would then be if any of the lower-dimensional
modes had a complex frequency, since the imaginary part would be associated
with a fluctuation that could grow exponentially in time.

   If the lower-dimensional background were just a Minkowski spacetime then
the criterion for classical linearised stability would simply be that
the masses of all the modes should be real.  However, because we are
concerned here with anti-de Sitter ground states, account must be taken of
the fact that it is now AdS representations, rather than Poincar\'e 
representations, that are relevant.  This was addressed in detail
in the work of Breitenlohner and Freedman \cite{breifree}.  The upshot is
that scalar fields in four-dimensional AdS spacetime can actually have
unitary representations (with real frequencies) even if the (mass)$^2$ is
negative, provided that it is not too negative.  Specifically, the
criterion is that if the mass obeys $M^2\ge - m^2$, where the cosmological
constant of the AdS$_4$ spacetime is given by $R_{\mu\nu}=\Lambda\, g_{\mu\nu}$
and $\Lambda=-12m^2$, then the representation will be unitary.  This is
the so-called Breitenlohner-Freedman (BF) bound.

   A detailed analysis of the mass spectrum for an arbitrary Freund-Rubin
compactification of eleven-dimensional supergravity was described in
\cite{dunipo}, with the mass spectra of the various four-dimensional fields
given in terms of the eigenvalue spectra of certain differential 
operators on the internal seven-dimensional compactifying space.  It turns
out that there is only one tower of four-dimensional modes that is at risk
of giving rise to classical instabilities \cite{Duff:1984sv}, 
and that is one of the
scalar field towers for which the masses are given by
%%%%%
\be
M^2 = \Delta_L - 4 m^2\,,
\ee
%%%%%
where $\Delta_L$ is the Lichnerowicz operator that acts on 
transverse, tracefree symmetric tensors $Y_{mn}$ on the internal 
seven-dimensional manifold:
%%%%%
\be
\Delta_L\,Y_{mn}= - \square Y_{mn} - 2 R_{mpnq}\, Y^{pq} +
  2 R_{(m}{}^p\, Y_{n)\, p}\,,\quad \nabla^m\,Y_{mn}=0\,,\quad
g^{mn}\, Y_{mn}=0\,.\label{Lichdef}
\ee
%%%%%%%
 Thus the criterion for classical linearised stability is that the
eigenvalues of the Lichnerowicz operator should be such that 
\cite{Duff:1984sv} 
%%%%%%%%
\be
\Delta_L \ge 3 m^2\,.\label{BFstab}
\ee
%%%%%

    Establishing the classical stability of a given Freund-Rubin
background thus reduces to the problem of finding the lower bound
on the eigenvalues of the Lichnerowicz operator on the
compact Einstein seven-manifold $X_7$. 

   It can be shown straightforwardly that any $X_7$ that admits 
Killing spinors will be such that $\Delta_L \ge 3m^2$ \cite{Duff:1984sv}, 
and thus all supersymmetric compactifications will be classically stable.
This accords with the usual expectation that supersymmetry protects
solutions against instabilities.  It is also noteworthy that for any
$X_7$ admitting fewer than the maximal ${\cal N}=8$ Killing spinors
of the round $S^7$, then if the orientation of the manifold is reversed
the resulting compactification will admit no supersymmetries.  Thus
classically, these ``skew-whiffed'' non-supersymmetric compactifications 
will also be classically stable \cite{dunipo,Duff:1984sv}.

   It can easily be seen that if $X_7$
is a direct product of two Einstein manifolds, $X_7=X_{(1)}\times X_{(2)}$,
where the two product manifolds have dimensions $N_1$ and $N_2$,
then an unstable mode can constructed very simply, by taking $Y_{mn}$
to be of the form $Y_{mn}= \hbox{diag}\, (\epsilon_1\, g^{(1)}_{m_1n_1},
\epsilon_2\,g^{(2)}_{m_2 n_2})$, where $\epsilon_1$ and $\epsilon_2$
are constants such that $N_1\,\epsilon_1 + N_2\, \epsilon_2=0$.  This
mode describes a perturbation in which one factor in the product expands 
uniformly while the other contracts, keeping the overall volume of $X_7$
fixed.  It can easily be seen that it has eigenvalue $\Delta_L=0$,
and hence it violates the BF stability condition (\ref{BFstab}).

   In general, the investigation of the classical stability of a
given Freund-Rubin compactification requires a detailed study of the
spectrum of the Lichnerowicz operator on $X_7$.  This was carried out for
various classes of coset-space examples in \cite{pagpop1,pagpop2,pagpop3}.
The technique employed in these papers was to show by an integration by 
parts on the compact Einstein manifold $X_7$ that
%%%%%
\be
\int_{X_7} Y^{mn}\, \Delta_L\, Y_{mn}\, dV =
\int_{X_7} \Big( -4 Y^{mn}\, R_{mpnq}\,Y^{pq} + 24 m^2\, Y^{mn}\, Y_{mn}
   +3 (\nabla^m\, Y^{np}\, \nabla_m\, Y_{np})\Big)\, dV\,.
\ee
%%%%%
Since the last term on the right-hand side is manifestly non-negative,
this means that the Lichnerowicz operator can be bounded from below
in terms of the upper bound on the eigenvalues of the Riemann tensor.
That is to say, the Riemann tensor has 27 eigenvalues $\kappa$,
defined by
%%%%%
\be
R_{mpnq}\, X^{pq}= \kappa\, X_{mn}\,,
\ee
%%%%%
where $X_{mn}$ denotes a symmetric eigentensor. 
Because the $X_7$ manifolds considered in \cite{pagpop1,pagpop2,pagpop3}
were all coset spaces, the eigenvalues $\kappa$ were all simply constants.
A criterion 
(\ref{BFstab}) for classical vacuum stability will then be satisfied for
all the modes if the condition
%%%%%
\be
\kappa_{\rm max} \le \frac{21}{4}\, m^2\,,\label{Riemcon}
\ee
%%%%%
holds,
where $\kappa_{\rm max}$ denotes the largest eigenvalue of the Riemann 
tensor.  Since the Riemann tensor can be calculated explicitly for
all the coset space examples, this means that the eigenvalues can be
obtained explicitly too.  Note that the inequality (\ref{Riemcon}) is
a {\it sufficient} condition for classical stability, but it is 
not necessarily {\it necessary}. 

   As an example, consider the class of $M^{mn}$ coset spaces that
were discussed in \cite{pagpop1}, which can be thought of as taking 
the form $SU(3)\times SU(2)/(SU(2)\times U(1))$, with the
integers $m$ and $n$ characterising the embedding of the denominator
group in the numerator. If $m=3$ and $m=2$ there exist two Killing spinors,
and the vacuum has ${\cal N}=2$ supersymmetry; this case, of course, will
necessarily be classically stable.  There exists an Einstein metric on
$M^{mn}$ for all integers $m$ and $n$.  It was shown in \cite{pagpop1} that
of the 27 eigenvalues of the Riemann tensor, 26 always satisfy the
stability bound $\kappa\le\ft{21}{4}\, m^2$, and so the possibility for
a classical instability hinges upon the 27th eigenvalue, which is
$\kappa_{\rm max}$.  This turns out to violate the bound (\ref{Riemcon})
if the ratio $2m/(3n)$ lies sufficiently far away, in either direction, 
from the $2m/(3n)=1$ supersymmetric case.  This establishes that in
cases outside the range $C_{\rm min}\le 2m/(3n)\le C_{\rm max}$, the 
possibility of a classical instability is not ruled out.  It was
furthermore then shown in \cite{pagpop1}, by explicit construction, that
a mode that violates the bound (\ref{Riemcon})
does in fact exist if $2m/(3n)<C_{\rm min}$ or $2m/(3n)> C_{\rm max}$.
Thus putting these results together, it was shown that the $M^{mn}$ 
Freund-Rubin compactifications are classically stable if 
$C_{\rm min}\le (2m/(3n)\le C_{\rm max}$, and that they are classically 
unstable otherwise \cite{pagpop1}.

   A much more difficult case to analyse is if one is considering 
an inhomogeneous metric on the compact manifold $X_7$.  One could,
for example, consider inhomogeneous Einstein metrics
on $S^7$.  As was shown by Bohm \cite{Bohm}, there exists a discrete
infinity of inequivalent Einstein metrics in this class.  There
exist no Killing spinors in any of the inhomogeneous
metrics, and so there is no particular reason to expect that the
corresponding Freund-Rubin vacuum solutions would be classically
stable.  The Bohm metrics are of cohomogeneity one, so the
analysis of the Lichnerowicz operator involves the study of
ordinary differential equations.  Some procedures for obtaining
numerical bounds on the eigenvalues of the Lichnerowicz operator
were discussed in \cite{gihapo}.

In the last decade, the issue of the 
stability of AdS compactifications has become one of the 
key questions addressed in 
 the Swampland project.  It has been conjectured by Ooguri and 
Vafa \cite{oogvaf} that
any non-supersymmetric compactification containing
an AdS factor must be unstable. Of the many such compactifications that have 
appeared over the years
a vast majority have already been proven unstable by appealing to 
various kinds of decay modes. As it turns out, the right-squashed $S^7$
is one of a very small number that has still not been proven to be unstable.
This remains an interesting topic for further research.

%%%%%%%%%%%%%%%%%%%%%%%%%%%%%%%%%%%%%%%%%%%
\section{Consistent truncations}

  Kaluza-Klein supergravity has evolved in many different ways in the 
period since the publication of our Physics Reports \cite{dunipo}. Some
of these have been as a consequence of totally unexpected developments,
of which perhaps the most surprising was the AdS/CFT correspondence, which
was first conjectured in 1997 \cite{malda,gkp,wit}. 

   In the years following the
writing of our Physics Reports the notion that the four-dimensional
theories obtained by compactifying $D=11$ supergravity on a compact coset
space such as a sphere might directly be interpretable as candidate 
theories of the real world was becoming less and less tenable.  For one thing,
it would be difficult, to say the least, to reconcile the concomitant 
huge negative cosmological constant with our everyday experience and with
astronomical observations.  It had also been shown by Witten that there
would never be any hope of obtaining a four-dimensional theory with 
a chiral fermionic sector, which would be more or less a {\it sine qua non} for
any realistic theory of the world \cite{Wittensearch}.  At the same
time, dramatic progress was being made in string theory, with the floodgates
opening on the development of quasi-realistic four-dimensional models,
following the breakthrough paper of Candelas, Horowitz, Strominger and
Witten on the compactification of the heterotic string on
Calabi-Yau 3-folds \cite{cahostwi}.

   Impressive and elegant though the new string theory results were, this new
direction in the quest for a unifying theory was perhaps a little
disappointing for those who had been working previously in Kaluza-Klein
supergravity.  One of the most appealing aspects 
of the traditional Kaluza-Klein 
approach, which had attracted many researchers, was the idea that the
local symmetries of the gauge fields in the lower dimensional theory 
found their origin in the geometrical symmetries (isometries) of the
compactifying space.  By contrast, in the Calabi-Yau compactifications of
the heterotic string the internal Calabi-Yau manifold has no continuous
symmetries at all, and the gauge fields that are present in the lower
dimension are more or less the same, or a subset, of those that were already
present in ten dimensions.

  It was at this point, in 1997, that the AdS/CFT correspondence burst onto
the scene. Suddenly, supergravity compactifications with AdS vacua were
fashionable again.  All the technology that had been developed in the
halcyon days of the coset compactifications had an application again. No
longer, admittedly, in the role of providing (or failing to provide) 
realistic four-dimensional
unified theories of the fundamental forces, but as the key player on one
side of the new duality symmetries, relating properties of classical
fields in AdS supergravity in $D+1$ dimensions to the quantum properties of
operators in $D$-dimensional boundary conformal field theories at 
strong coupling.

   As well as providing a new arena in which the coset compactifications
of supergravities had a leading role, the AdS/CFT correspondence gave
a new stimulus to another aspect of the coset compactifications, which
had long been a source of fascination to those who had played around
with the Kaluza-Klein reductions, but which had perhaps always seemed a 
bit like a solution in search of a problem.  The object of this
fascination is the notion of consistent reductions.

   The earliest Kaluza-Klein compactifications that were considered involved
just reducing a higher-dimensional theory on a circle.  The idea then is that
one assumes all the higher-dimensional fields are independent of the 
circle coordinate, say $z$.  One can think of this as first performing 
Fourier expansions of all the higher-dimensional fields with respect to the
$z$ coordinate, and then retaining only the zero mode in each of the
Fourier expansions.  The resulting lower-dimensional theory is then guaranteed
to be a consistent truncation of the complete lower-dimensional theory,
comprising infinitely-many fields, that would have been obtained if
all the infinite towers of Fourier modes had been retained.  The notion
of consistency here means that any solution of the truncated 
lower-dimensional theory will also be a solution of the full untruncated
theory.  To put it another way, any solution of the lower-dimensional 
truncated theory will lift, by inverting the Kaluza-Klein reduction step, 
to give a solution of the original higher-dimensional theory.  

   One way of stating the consistency of the truncation is that prior
to truncation, the full lower-dimensional equations of motion for
the untruncated non-zero mode fields will be such that it is consistent
to set all the non-zero mode fields to zero. That is to say, there
cannot be any source terms in the full non-zero mode equations that are
formed purely from the zero-mode fields that are eventually to be retained.
It is easy to see in this example why this is guaranteed to be true.  Namely,
the non-zero mode fields are all {\it charged} under the $U(1)$ symmetry
group that acts on the circle, while the zero-mode terms are all 
{\it uncharged}.  No matter how non-linear the equations may be, it is 
impossible for powers of purely uncharged fields to generate a term that
could act as a source for a charged field.  Put yet another way, the
consistency is guaranteed because one is truncating to the {\it singlet sector}
of the $U(1)$ group that acts on the circle.\footnote{Of course, for the
argument of consistency to work, it is essential to retain {\it all} the
singlet fields in the truncation.  An example where consistency in a circle
reduction would fail is if one reduced five-dimensional pure gravity on
a circle, and failed to retain all the $U(1)$ singlets in the truncated
four-dimensional theory.  The full set of singlets comprises the
four-dimensional metric, the Kaluza-Klein vector originating from
the mixed $(\mu z)$ components of the higher-dimensional metric
$\hat g_{MN}$, and the dilatonic scalar originating from the $(zz)$
component of $\hat g_{MN}$.  In some early attempts at Kaluza-Klein reduction
the dilatonic scalar was erroneously truncated out, seemingly giving
a nice Einstein-Maxwell theory in four dimensions.  However, the solutions
of Einstein-Maxwell do not lift back to solutions of the five-dimensional
Einstein theory, because the Maxwell field would actually act as a source,
via an $F^{\mu\nu}\, F_{\mu\nu}$ term, for the dilaton that was supposedly 
set to zero.}

   The consistency of the truncation in the circle reduction generalises
immediately to the multi-step reduction on a number of circles; that is,
a reduction on a torus.  It also straightforwardly generalises to
another class of reductions, where the compactifying space is taken to
be a group manifold $G$.  The manifold $G$ admits a bi-invariant metric,
invariant under independent left and right actions of the group $G$. Each
of these groups $G_L$ and $G_R$ acts transitively on $G$.  As was observed
by Bryce DeWitt in 1963, one can write down a reduction ansatz for 
a higher-dimensional metric in which the gauge bosons of just one of these
symmetry factors, say $G_R$, are retained \cite{DeWitt}.  This is guaranteed
to be a consistent reduction, because the fields that are being retained
comprise all the singlets under the action of the group $G_L$, while all
the fields that have been truncated are non-singlets under $G_L$.  The
consistency follows because powers of $G_L$ singlets cannot act as sources for $G_L$ non-singlets.  Because $G_L$ acts transitively, there will be just
a finite number of fields in the truncated subset.  Consistent reductions
of this kind are sometimes referred to as DeWitt reductions.

  The consistency of a reduction on a general coset space $G/H$ is a totally
different story.  First, we should clarify that the truncations we have in
mind in all these cases are ones in which one retains, among other fields,
the gauge bosons of the isometry group $G$ of the coset.  In other words,
we are interested in consistent truncations where the original Kaluza-Klein
dream of obtaining the gauge symmetries of the lower-dimensional fields from
the geometric isometry symmetries of the compactifying manifold is
realised.  This was
an idea that had occurred to Pauli in 1953.  He wrote a letter to A. Pais,
proposing the idea that one might reduce six-dimensional Einstein gravity
on a 2-sphere, thereby obtaining a four-dimensional theory with $SU(2)$
Yang-Mills fields (they weren't called Yang-Mills fields in those days). 
But he also killed off his own idea in the same letter, because
he had appreciated that it would not be a consistent reduction.  
(An account of this early proposal for a non-abelian Kaluza-Klein 
reduction can be found in \cite{straum,orafstra}.)
In terms
of the discussion given above, if one expanded the components of the 
six-dimensional metric in terms of spherical harmonics on the 2-sphere, then
one would find that the non-zero modes in the expansion of the four-dimensional
metric components (i.e. spin-2 tensor fields in four dimensions) 
 would have source terms built from the $SU(2)$ Yang-Mills fields one wishes
to retain.  Thus it would be inconsistent to set the massive spin-2 fields 
to zero.  

    Blissfully ignorant of the potential pitfalls ahead, in the early
1980s we and others in the supergravity community began to explore the
idea of obtaining four-dimensional $SO(8)$-gauged ${\cal N}=8$ supergravity
by compactifying $D=11$ supergravity on the 7-sphere.  All the portents 
in a linearised analysis looked 
good; the sphere has an $SO(8)$ isometry, and the Freund-Rubin 
AS$_4\times S^7$ vacuum solution has ${\cal N}=8$ supersymmetry \cite{dufpopKK}. 

    The first
person to rain on the parade was Gary Gibbons, who simply 
said ``it won't work.''  He even went so far as to declare that he would
``eat his hat'' if it worked.  When we finally understood what he was talking
about, we realised that indeed he had a point; this was seemingly a situation
very like the one Pauli had encountered thirty years before (although we didn't
know about Pauli's attempts back then).  Indeed, it looked dangerously
likely that the truncation would not be consistent beyond the linearised level;
the $SO(8)$ Yang-Mills fields would seemingly act as sources for massive
spin-2 fields in the four-dimensional Kaluza-Klein towers.  We persevered
anyway, somehow convincing ourselves that ``the 7-sphere knows what it's 
doing,'' and that something would emerge to save the day.  And indeed,
not long after, we found a glimmer of hope in a calculation we did with
Nick Warner \cite{dunipowa}.  We examined the expected leading-order
stumbling block to the consistency of the truncation, namely the
term where quadratic products of $SO(8)$ Yang-Mills fields would
act as sources for massive spin-2 fields.  Remarkably, it turned out 
that the coefficient of this dangerous coupling was exactly zero,
by virtue of a seemingly miraculous conspiracy between two separate
contributions, one being from the metric contribution in the Einstein
equations (just as Pauli would have had), and the other being a 
contribution from the $SO(8)$ gauge fields in the reduction ansatz for
the eleven-dimensional 4-form field strength and its energy-momentum tensor.
So indeed it seemed the 7-sphere ``knew what it was doing,'' but in a 
rather subtle way that involved a conspiracy with the $D=11$ supergravity
theory as well.  It gave us hope that the miracles would continue, and that
the consistency of the truncation would persist to all orders.

  Not long after that, in a true {\it tour de force}, Bernard de Wit and 
Hermann Nicolai constructed a complete metric reduction ansatz, and 
provided a demonstration of the full consistency of the 7-sphere truncation 
\cite{dewnic} (a few loose ends involving the explicit forms of some of 
the components of the 4-form reduction ansatz were only sorted out a while
later; see, for example, \cite{NP,godgodnic}).  

   To those who had been involved in the efforts to demonstrate the
consistency of the reduction, it was almost magical the way the pieces
of the puzzle fitted together, and the whole picture was one of great
beauty.  Not everyone shared the wonderment, however, and it was
not uncommon in those days to face scepticism, or worse, from audience
members at a seminar whose reaction was along the lines of ``why should
we care about consistency, we are only interested in the low-energy
effective theory anyway.''

  Undeterred, others who appreciated the elegance of the Pauli type of 
consistent reductions continued investigating other instances where they
can arise.  A nice example was obtained by 
Horatiu Nastase, Diana Vaman and Peter van Nieuwenhuizen, where they
constructed the consistent reduction of $D=11$ supergravity on
the 4-sphere, thereby arriving at maximal gauged supergravity in
seven dimensions \cite{nasvamvan1,nasvamvan2}.  This example is actually
rather simpler than the 7-sphere reduction, largely because the
scalar field sector of the reduction is easier to handle.\footnote{It is
always the scalar sector that creates the biggest headaches in the
construction of consistent truncations.}  In fact, the 4-sphere
reduction pointed the way to various other examples of 
consistent reductions, including the $SL(2,{\mathbf R})$-singlet
subsector of the consistent truncation
for the 5-sphere reduction of type IIB supergravity \cite{cvluposatr}.
The construction of the full consistent truncation of type IIB supergravity
on $S^5$, to give maximal gauged supergravity in five dimensions, 
proved to be more recalcitrant, owing once again to a rather tricky scalar
sector, and it was only following another crucial development, to be
chronicled below, that this one was finally achieved.

   One nice footnote to these developments takes us forward to 2006,
when a conference in honour of Gary Gibbons' 60th birthday took place 
in Cambridge.  Hermann Nicolai had secretly arranged with Fitzbillies
in Cambridge to create a cake in the shape of a hat, and at the end of
his talk at the conference Hermann presented Gary with the cake and 
invited him to eat his hat.  Very gamely, Gary proceeded to consume
it in front of the conference audience, thus fulfilling a promise he
had made more than twenty years before.

Things began to change, in some quarters at least, with the developments
in the AdS/CFT correspondence.  Now, properties of the AdS field theories
that arose from the sphere reductions of higher-dimensional supergravities 
translated into properties of the amplitudes for quantum operators on the
CFT side of the duality.  It could be very helpful to know whether
or not one was dealing with a properly self-contained subset of the 
complete towers of lower-dimensional fields in the AdS background.  Also, the
existence of a consistent truncation meant that one could sometimes exploit
it in order to construct exact solutions in the higher dimension by
starting with more easily constructed exact solutions in the lower dimension,
and then lifting them up by using the consistent reduction ansatz in reverse.

  One important application of the AdS/CFT correspondence is the ABJM
theory \cite{abjm}, which provides a holographic dual of M-theory 
compactified on AdS$_4\times S^7/{\bf Z}_k$, thus providing a holographic description of
the 3-dimensional world-volume theory of M2-branes.  The near-horizon geometry of the M2-brane
is AdS$_4\times S^7 $ \cite{Duff:1990xz}.
In fact some of the ideas employed
in the ABJM theory were foreshadowed by a paper written almost 25 years
previously \cite{nilpophopf}, where it was observed that since 
the 7-sphere can be viewed as a $U(1)$ bundle over $\CP^3$ , one could
make a Kaluza-Klein reduction of the $S^7$ compactification of $D=11$
supergravity on the Hopf circle, and interpret it as compactification 
of $D=10$ type IIA supergravity on $\CP^3=SU(4)/U(3)$.  Keeping only the $U(1)$
singlets in the usual fashion, this means that only an ${\cal N}=6$
supersymmetry survives, since the 8 of $SO(8)$ gravitini from $S^7$ 
decompose as a $6_0+ 1_2+1_{-2}$ under $SU(4)\times U(1)$.  Of course,
the truncation is guaranteed to be a consistent one.

   The next major development in the story of consistent reductions 
takes us into the realm of generalised geometry and exceptional field
theory (ExFT).  Perhaps one of the earliest precursors of these ideas goes
back to another paper by de Wit and Nicolai, in 1985 \cite{dewnic2},
where they showed how $D=11$ supergravity could be written 
in terms of four-dimensional fields in a manner
suggestive of the way one would carry out a consistent truncation on
the 7-torus, but without actually making any truncation of the fields. This
new formulation of the $D=11$ theory has a local $SU(8)$ invariance,
with the bosonic quantities relating to the scalar fields carrying 
56 and 133 dimensional representations of the $E_{7(7)}$ 
Cremmer-Julia global symmetry of the usual toroidally-compactified theory.
The idea of adding extra coordinates in the description of 
duality symmetries in membrane theory was also introduced, in  
\cite{dufflu}. 

   The notions of generalised geometry and exceptional field theory
push the ideas of de Wit and Nicolai further.  In order to exploit the
symmetries further, one now introduces additional coordinates, in order
to reformulate the supergravity theory in such a way that the
$E_{7(7)}$ Cremmer-Julia global symmetry of the 7-torus truncation is
now realised as a symmetry of the extended theory without any truncation
being performed.  Geometric structures, such as a generalised
Lie derivative, are then defined on the extended space. The corresponding 
algebra of generalised diffeomorphisms closes, provided that one imposes
a so-called section condition.  The section condition itself is written
in an $E_{7(7)}$-covariant way. 
Any specific solution of the section 
condition amounts to a restriction in which the fields are allowed to
depend only on a subset of the extended system of coordinates, and 
correspondingly the $E_{7(7)}$ symmetry is broken.\footnote{This feature
of generalised geometry has been known to give conference speakers an
uncomfortable time in the question session.  However, it is perhaps
rather reminiscent of the situation in gauge theory, where the gauge
symmetry is introduced, but is then eventually lost again when restricting
to the sector of physical states.}  However, since
the formulation of the extended theory itself is $E_{7(7)}$ covariant, 
prior to choosing a specific solution to the section condition, this provides
a framework within which the $E_{7(7)}$ symmetry can be employed in order
to construct truncations whose consistency can be understood rather
straightforwardly. 
In fact it turns out that the formalism 
can provide an understanding of why a non-trivial Pauli type of
sphere reduction can actually yield a consistent truncation.  Essentially,
it can be recast as a statement of truncation to fields that are singlets
under the enlarged symmetry group of the extended formulation, thereby making
the consistency of the truncation manifest.

  As stated above, the construction was rather specific to reformulating 
$D=11$ supergravity in an extended $(4+7)$-dimensional language.  Similar
reformulations in an $([11-n]+n)$-dimensional language are also
possible for a range of $n$, corresponding to the cases where the there
would be an $E_{n(n)}$ global symmetry in a consistent truncation on the
$n$-torus.  An analogous set of generalisations of type IIB supergravity,
viewed from a $([10-n]+n)$-dimensional standpoint, are also possible.
One of these, namely the $(5+5)$ generalisation, was employed a few 
years ago to realise the long-standing goal of constructing the 
consistent truncation of type IIB supergravity on $S^5$, yielding 
maximal gauged five-dimensional supergravity \cite{baghohsam}.   
In another notable paper, the methods of generalised geometry and exceptional
field theory were used in order to revisit the consistent truncation of
$D=11$ supergravity on the 7-sphere \cite{varela}.

  The ExFT methods that were used in order to construct the consistent 
reduction of $D=11$ supergravity on $S^7$ to give the ${\cal N}=8$ maximal
gauged supergravity in four dimensions were subsequently generalised in
order to find the Kaluza-Klein spectra for any vacuum that
can be described as a deformation of the ${\cal N}=8$ vacuum that is 
implemented by fields living in the ${\cal N}=8$ 
supermultiplet \cite{malsam1,malsam2}.  This includes the wide class of 
known extrema of the ${\cal N}=8$ supergravity potential. Not included
in this class, however, is the ${\cal N}=1$ squashed $S^7$ vacuum, since
in this case it corresponds to a deformation of the round $S^7$ vacuum by
fields that lie outside those of ${\cal N}=8$ supergravity (see, for example,
\cite{dunipo}).  Recently, even for this case the Kaluza-Klein spectrum 
was constructed using ExFT techniques \cite{dubmalsam}.  The results
are compatible with those obtained recently in \cite{Karlsson:2023dnl}, 
where the full Kaluza-Klein spectra in the ${\cal N}=1$ and
${\cal N}=0$ squashed $S^7$ vacua were obtained by using coset-space harmonic
analysis.

%%%%%%%%%%%%%%%%%%%%%%%%%%%%%%%%%%%%%%%%%%%%%%%%
\section{The complete form of the squashed $S^7$ spectrum}
%%%%%%%%%%%%%%%%%%%%%%%%%%%%%%%%%%%%%%%%%%%%%%%%
In this section we  first present some of the more recent developments concerning the improved techniques used to solve
eigenvalue equations on complicated coset spaces like the squashed $S^7$ \cite{Karlsson:2021oxd, Karlsson:2023dnl}.  We then apply it to the squashed $S^7$ and give a brief account of  the key results that have been obtained
this way. The explicit construction of the 2-form mode functions in \cite{Karlsson:2023dnl} is a novel result in this context which 
used ideas that appeared first in \cite{Nilsson:1983ru}. With these eigenmodes at hand their  eigenvalues can be computed which makes it possible to obtain a detailed understanding of the entire ${\cal N}=1$ supermultiplet spectrum for the left squashed $S^7$ compactification.These new developments that started in 2018 with the derivation of the complete isometry irrep spectrum \cite{Nilsson:2018lof} fill in more or less all 
gaps that were left when the subject
came to a halt at the end of the 1980s. At that point only the spectra of $\Delta_0$, $ i\slashed D_{1/2}$ 
\cite{Nilsson:1983ru} and $\Delta_1$ \cite{Yamagishi:1983ri} were understood in detail. In addition the eigenvalues of the Lichnerowicz operator $\Delta_L$ were quoted in \cite{dunipo} without proof. Some of the issues that still remain to be clarified are mentioned at the end of this section.

%%%%%%%%%%%%%%%%%%%%%%%%%%%%%%%%%%%%%%%%%%%%%%%%%%
\subsection{Improved formalism on general reductive coset spaces $G/H$ } 
%%%%%%%%%%%%%%%%%%%%%%%%%%%%%%%%%%%%%%%%%%%%%%%%%%
The eigenvalue equations that we need to solve on any internal manifold $K^7$ 
used in a Freund-Rubin compactification of $D=11$ supergravity in order 
to find the mass spectrum in the AdS$_4$ supergravity theory involve the following operators:
\begin{equation}
\Delta_p=\delta d+d \delta \,\,(p=0,1,2,3), \,\,\,\Delta_L, \,\,\,Q=\star d,\,\,\, i\slashed D_{1/2}, \,\,\,i\slashed D_{3/2}, 
\end{equation}
where $\Delta_3=Q^2$ and $\Delta_L$ is the Lichnerowicz operator which arises in the variation of the Ricci tensor.
By squaring also the two spinorial operators in this list we get a set of operators that can be collectively be written  in terms of the Riemann tensor as \cite{Christensen:1978md, Karlsson:2021oxd}
\begin{equation}
\Delta=-\Box-R_{abcd}\Sigma^{ab}\Sigma^{cd},
\end{equation}
where $\Sigma^{ab}$ are the generators of the tangent space group $SO(7)$. This form is a rather trivial but extremely useful rewriting of the standard expressions for these
operators.

A convenient  form of the Riemann tensor in this context is \cite{Bais:1983wc}
\begin{equation}
R_{abc}{}^{d}=f_{ab}{}^if_{ic}{}^d+\frac{1}{2}f_{ab}{}^ef_{ec}{}^d+\frac{1}{2}f_{[a|c|}{}^ef_{b]e}{}^d.
\end{equation}
Here the structure constants are the ones for a reductive coset manifold $G/H$ where the generators $T_A$ of the isometry group $G$
are split into $T_i$ of the  subgroup $H$  and the rest $T_a$. The non-zero commutators are thus
\begin{equation}
[T_i, T_j]=f_{ij}{}^kT_k,\,\,\,\, [T_i, T_a]=f_{ia}{}^bT_b,\,\,\,\,[T_a, T_b]=f_{ab}{}^i\,T_i+f_{ab}{}^c\,T_c.
\end{equation}

In the  application of these equations to non-trivial coset spaces like the one of the squashed $S^7$, that is   $(Sp_2\times Sp_1^C)/(Sp_1^A\times Sp_1^{B+C})$\footnote{The factor $Sp_1^{B+C}$ is the diagonal subgroup of $Sp_1^C$ and $Sp_1^B$ from the $Sp_2$ subgroup $Sp_1^A \times Sp_1^B$.}, there is  an extremely useful relation that 
expresses an $H$-covariant derivative algebraically \cite{Salam:1981xd, dunipo}:\\
\be
\check D_a \equiv D_a+\frac{1}{2}f_{abc}\Sigma^{bc}=-T_a
\ee
 where $D_a$ is the covariant derivative on the tangent space of $G/H$, with dimension $n=dim(G)-dim(H)$,  
and $\Sigma^{ab}$  are the $SO(n)$ generators. This implies that
\begin{equation}
\check\square\equiv \check D_a\check D^a= T_aT^a=-(C_G-C_H),
\end{equation}
where $C$ is a Casimir operator.  There are at least two aspects of this equation that we need to address. First, it is not $\check\square$ that appears in the eigenvalue equations we are interested in
but rather $\square$.  It is however a trivial exercise to derive 
a formula relating $\check\square$ to the $\square$. It reads
\begin{equation}
\check\square=\square+f_{abc}\Sigma^{ab}\check{D}^c-\frac{1}{4}f_{abc}f^{ade}\Sigma^{bc}\Sigma_{de}.
\end{equation}

The second aspect is that  the spectrum is supposed to depend only on the Casimir of the isometry group  $G$ and not of the isotropy group $H$.
However, using the above  equations we find the following general form of the universal 
Laplacian \cite{Karlsson:2021oxd, Karlsson:2023dnl}
\begin{equation}
\label{univlapl}
\Delta=C_G+f_{abc}\Sigma^{ab}\check{D}^c-\frac{1}{4}(3f_{abc}f^{a}{}_{de}-2f_{abd}f^{a}{}_{ce}\Sigma^{bc}\Sigma_{de}).
\end{equation}
We see here that the Casimir for the isotropy group $H$ has cancelled out which is a necessary feature for this to work. \\%After all we expect that the spectra
%will depend  only on the Casimir of $G$, and not on $H$.\\

Now the crucial  issue is how one is supposed to use the equations above. There are  two possible ways to proceed:\\
a) One can start squaring $f_{abc}\Sigma^{ab}\check{D}^c$ in (\ref{univlapl}) in order to find  non-linear equations for it which hopefully can be solved. In this case
one needs also the Ricci identity for $\check{D}_a$ which in terms of the $H$ generators $T^i$  reads 
\beq
[\check{D}_a, \check{D}_b]=(T^i)_{ab}(T_i)_{cd}\Sigma^{cd}+f_{ab}{}^c\check D_c.
\end{equation}
This approach will produce eigenvalues with no reference to the actual eigenmodes and hence one will have a rather poor understanding of the structure of the spectrum.\\
b) A much more powerful way to use the above equations would be to first explicitly construct all the eigenmodes and then hit them with $f_{abc}\Sigma^{ab}\check{D}^c$.
This will give rise to matrix equations that are quite complicated but can  be solved using some algebraic computer program.\\
c) A final point concerns the probable proliferation of $H$ irreps to study if the analysis cannot take advantage of some more structure in the equations. The reason why some additional structure 
may be expected is that it is known to occur in the case of the squashed seven-sphere. In that case the extra structure is connected to the holonomy group which is $G_2$.
In this particular case we have the sequence of groups \cite{dunipo}
\beq
SO(7)\rightarrow G_2 \rightarrow H=Sp_2\times Sp_1^C.
\end{equation}
As we will see in the next subsection the use of the structure connected to the group $G_2$ when solving the eigenvalue equations will 
simplify the procedure enormously. This will be particularly clear when the explicit 2-form mode functions are presented below.
The group $G_2$ will also give us the opportunity to make use of octonions. It would be interesting to see if the use of the  holonomy group $G_2$ 
and octonions has a counterpart in other examples of Kaluza-Klein compactifications.

 %%%%%%%%%%%%%%%%%%%%%%%%%%%%%%%%%%%%%%%%%%%%%%%%%%
\subsection{The full spectrum on the squashed seven-sphere}
%%%%%%%%%%%%%%%%%%%%%%%%%%%%%%%%%%%%%%%%%%%%%%%%%%

We now apply the above formalism to the case of the squashed seven-sphere\footnote{The full spectrum of the round seven sphere may be found in \cite{Englert:1983rn,Sezgin:1983ik,Biran:1983iy}.}. The extra information we then have is the explicit
expressions for the structure constants. In particular we find a direct connection to the octonions via
\beq
f_{abc}=-\frac{1}{\sqrt 5}a_{abc}
\end{equation}
where $a_{abc}$ are the octonionic structure constants \cite{Bais:1983wc}.  
This leads to the following property of the $H$-covariant derivative $\check{D}_a$ \cite{Ekhammar:2021gsg}
\beq
\check{D}_a\, a_{bcd}=0.
\end{equation}

When looking for supersymmetry in the squashed $S^7$ vacuum one solves the Killing spinor equation which involves a covariant derivative with $G_2$ holonomy.
This derivative is just $\check D_a$ when acting on the single solution to the Killing spinor equation $\eta$. 
The perhaps surprising relation above to the octonionic structure constants is in fact rather natural in view of the following connection to the $D=7$ gamma matrices
\beq
a_{abc}=i\bar\eta\Gamma_{abc}\eta.
\end{equation}

By defining some relevant  Casimir operators in terms of the structure constants $f_{AB}{}^C$ and using the fact that the definition of $\Delta$ above involves  two $SO(7)$ generators
one finds a remarkably simple and general formula for $\Delta$ \cite{Karlsson:2023dnl}:
\begin{equation}
\Delta=C_G+\frac{6}{5}C_{SO(7)}-\frac{3}{2}C_{G_2}-\frac{1}{\sqrt 5}a_{abc}\Sigma^{bc}\check D^a.
\end{equation}
Thus it has now become obvious that  for the eigenvalue equations and the mode functions the only relevant tensor properties are the ones of the  tangent space group $SO(7)$ and the holonomy group  $G_2$, in addition to those of 
the isometry group $G$ of course. 

As an example we apply the above $\Delta$ to the 2-form mode functions $Y_{ab}$, see  \cite{Karlsson:2021oxd} for the details. The 2-form eigenvalue equation in terms of $\Delta$ then becomes
\beq
\Delta_2 Y_{ab}=(C_G+3P_7-\frac{2}{\sqrt 5}\check D_{[2]})Y_{ab}=\kappa_2^2 Y_{ab},
\end{equation}
where we have defined $\check D_{[2]}\equiv a_{[a}{}^{cd}\check D_{|d}Y_{c|b]}$. To arrive at the above form of the eigenvalue equation we have used
the fact that under $SO(7)\rightarrow G_2$ we have ${\bf 21}\rightarrow {\bf 14}\oplus{\bf 7}$ and then used $C_{SO(7)}({\bf 21})=5$, $C_{G_2}({\bf 7})=2$ and $C_{G_2}({\bf 14})=4$. 
This also explains the presence of the $G_2$ projector $(P_7)_{ab}{}^{cd}=\frac{1}{6}a_{abe}a^{cde}$ in the equation above.
As explained in  \cite{Karlsson:2021oxd}, and in fact even earlier in  \cite{Ekhammar:2021gsg}, splitting the  2-form $Y_{ab}$ and the eigenvalue equation into their
 $\bf 7$  and $\bf 14$  pieces a short calculating, based on squaring $\check D_{[2]}$,  gives the result (using $\Delta_2$ instead of $\kappa_2^2$ and $m^2=9/20$) 
 \be
\Delta_2^{(1)}= C_G+\frac{18}{5}=\frac{m^2}{9}(20C_G+72),
 \ee
  \be
\Delta_2^{(2\pm)}= C_G+\frac{11}{5}\pm\frac{2}{\sqrt 5}\sqrt{C_G+\frac{49}{20}}=\frac{m^2}{9}\big(20C_G+44\pm 4\sqrt{20C_G+49}\big).
 \ee
  \be
\Delta_2^{(3)}= C_G=\frac{m^2}{9}20C_G,
 \ee
 
 Although these results do not tell us which modes (i.e., isometry irreps) are associated with which eigenvalues  we can make the observation
 that the 5-fold degeneracy of some (many in fact) irreps are  not met by  the same number of independent eigenvalues. In fact, as we see above there are only four of these. Thus,
 some eigenmodes must have the same eigenvalues. This will be demonstrated in detail when we now turn to the eigenmode calculations.

Previous attempts to construct all the 21 two-form mode functions met which considerable complications which were
overcome  using the
insights of the key role played by $G_2$ and octonions \cite{Karlsson:2023dnl}. Another important feature that made this mode construction feasible was that the appearance of
a double derivative (see $ { \mathcal Y}^{(6)i}_{ab}$ below) could easily be verified using the improved formalism presented above.
A more fundamental argument for this double derivative mode will be presented in \cite{Karlsson:2025} which also contains  a streamlined explanation
of the whole approach used here.

Before we start discussing the structure of the mode functions we need to figure out which isometry irreps they will carry. This question can be answered in two ways,
either by breaking the $SO(8)$ irreps of all modes in the round sphere spectrum down to $Sp_2\times Sp_1^C$ irreps, or by constructing them directly in the squashed coset.
Both approaches were used in \cite{Nilsson:2018lof} and  the result was presented in terms of so called cross diagrams for the irreps $(p,q;r)$. It turns out that there is a small set of irreps that 
appear only in one of the two vacua seemingly making a one-to-one relation of the modes in the two vacua impossible. That this discrepancy is really there in the spectra is strongly supported by
the construction of all ${\mathcal N}=1$ supermultiplets in the squashed vacuum  \cite{Karlsson:2023dnl}.

The ingredients needed to obtain the explicit form of the mode functions are:\\
1. The octonionic structure constants $a_{abc}$ and their dual $c_{abcd}$.\\
2. The derivative $\check D_a$ satisfying $\check D_a a_{bcd}=0$.\\
3. The Killing vector of the isometry group $Sp_1^C$: $s^i=s^{ia}\partial_a$ satisfying 
$[s^i, s^j]=\epsilon^{ijk}s^k$, and  
 \cite{Nilsson:1983ru}
\be
\check D_a\, s^i_b=-3\epsilon^i{}_{jk}s_a^js_b^k.
\ee

It can be argued that the 21 independent mode functions  of $Y_{ab}$ must appear in irreps of the isometry group factor $Sp_1^C$ as the sum of one ${\bf 1}$,   five $ {\bf 3}$, and one ${\bf 5}$. This leads to the following mode operators,
which by letting  them act on scalar mode functions produces a complete set of two-form mode functions:

% \begin{gather}
%\label{eq:modes:2-form:building-modes:first}
 \ba
  { \mathcal Y}^{(1)}_{ab} &=& a_{ab}{}^c\check{D}_c, \,\,\,\,\,
%  %  \begin{alignedat}{2}
       { \mathcal Y}^{(2)i}_{ab} = a_{ab}{}^c s_c{}^i,\\
      { \mathcal Y}^{(3)i}_{ab} &=& \epsilon^i{}_{jk} s_a{}^j s_b{}^k,\,\,\,\,
       { \mathcal Y}^{(4)i}_{ab} = s_{[a}{}^i \check{D}_{b]},\\
        { \mathcal Y}^{(5)i}_{ab} &=& c_{ab}{}^{cd} s_c{}^i \check{D}_d,\,\,\,\,
        { \mathcal Y}^{(6)i}_{ab} = a_{[a|}{}^{cd} s_{c}{}^i \check{D}_{d|b]},\\
%  %  \end{alignedat}
%  %  \\[4pt]
   { \mathcal Y}^{(7)ij}_{ab} &=& s_{[a}{}^{\{i|} a_{b]}{}^{cd} s_c{}^{|j\}} \check{D}_d
%\label{eq:modes:2-form:building-modes:last}
%\end{gather}
 \ea
 
 One new feature  of these mode operators is the double derivative in $ { \mathcal Y}^{(6)i}_{ab}$. In the new improved formalism it is a fairly easy
 exercise to check that such an operator must be introduced. It is also a straightforward but rather lengthy computation, best done in Mathematica, 
 to obtain both the transverse modes and their eigenvalues. A detailed account of this can be found in \cite{Karlsson:2023dnl}.
 
 Once an exact relation between the $\Delta_2$ eigenvalues and the eigenmodes is established one confirms that
 there is a degeneracy in this spectrum. This feature permeates then  into parts of  spectrum of  ${\cal N}=1$ Heidenreich  supermultiplets \cite{Heidenreich:1982rz}
 as can be seen in the tables presented in \cite{Karlsson:2023dnl}. Interestingly enough, this seems to afflict only
 supermultiplets containing scalar fields which are tied to the Lichnerowicz operator.
 
 Another somewhat surprising property of the left squashed spectrum is that for one of the fourteen towers of  Wess-Zumino multiplets supersymmetry can be implemented in two different ways \cite{Karlsson:2023dnl}. This is made possible by choosing 
 different boundary conditions and hence different signs in the equations relating   $E_0$ to the mass of spin zero and 1/2 fields in AdS$_4$.

Furthermore, if one assumes that there is a one to one connection between the round
 and squashed spectra  it seems necessary to introduce singletons as part of the spectrum, as 
suggested in \cite{Nilsson:2018lof, Nilsson:2023ctq, Karlsson:2023dnl}.  Singletons were discovered by Dirac in 1963 \cite{Dirac:1963ta} and have been studied since the 1970s by Fronsdal and collaborators and many others. See,  for example \cite{Flato:1999yp,Starinets:1998dt,Nicolai:1984gb,Blencowe:1987bn, Samtleben:2024zoy}. They were discussed as a possible sector of the round $S^7$ spectrum in \cite{Sezgin:1983ik, Casher:1984ym, Sezgin:2020avr}.

The squashed $S^7$ vacuum supergroup $OSp(4,1)$ is not a subgroup of the round $OSp(4,8)$ since the massless gravitino is not one of the original 8. It is not obvious that it should obey the same rules as  
conventional Higgs and Superhiggs in  AdS$_4$ \footnote{See for example the decomposition rules for $SO(2,3)$ reps of Fronsdal \cite{Fronsdalsing} and  Porrati \cite{Porrati:2001db}.}. Their similarities and differences remain an interesting topic for future enquiry.

\section*{Acknowledgements}

MJD is grateful to Massimo Porrati and Leron Borsten for useful conversations and  STFC Consolidated Grants ST/T000791/1 and
ST/X000575 for support. The work of C.N.P. 
is supported in part by DOE grant DE-SC0010813.  
B.E.W.N is  grateful to Joel Karlsson for a very stimulating and productive 
collaboration that generated some of the results reviewed here. B.E.W.N. thanks Chalmers University of Technology, 
G\"oteborg, for the continued affiliation and partial funding of this work.
%\bibliographystyle{utphys}
%\bibliography{weyl.bib}

  \end{document}